\documentclass{amsart}
\usepackage{amsmath,amssymb}
\usepackage{graphicx,subfigure}
\usepackage[latin1]{inputenc}
\newtheorem{theorem}{Theorem}[section]

\newtheorem{lemma}[theorem]{Lemma}

\theoremstyle{definition}
\newtheorem{definition}[theorem]{Definition}
\newtheorem{example}[theorem]{Example}

\theoremstyle{remark}
\newtheorem{remark}[theorem]{Remark}

\numberwithin{equation}{section}

\newcommand{\al}{\alpha}
\newcommand{\be}{\beta}
\newcommand{\de}{\delta}
\newcommand{\ep}{\epsilon}

\newcommand{\ga}{\gamma}

\newcommand{\la}{\lambda}
\newcommand{\om}{\omega}
\newcommand{\si}{\sigma}

\newcommand{\De}{\Delta}
\newcommand{\Ga}{\Gamma}
\newcommand{\La}{\Lambda}
\newcommand{\Si}{\Sigma}
\newcommand{\Om}{\Omega}
\newcommand{\tPsi}{\widetilde{\Psi}}

\newcommand{\tL}{L}

\newcommand{\tv}{\tilde{v}}

\def\NN{\mathbb{N}}
\def\RR{\mathbb{R}}

\def\ZZ{\mathbb{Z}}

\renewcommand\SS{\mathbb{S}}
\newcommand{\cO}{{\mathcal O}}

\newcommand{\pd}{\partial}
\newcommand\minus\backslash
\newcommand{\id}{{\rm{id}}}

\newcommand{\ms}{\mspace{1mu}}
\newcommand\lan\langle
\newcommand\ran\rangle

\newcommand{\card}{\operatorname{card}}

\newcommand{\e}{{\mathrm e}}

\DeclareMathOperator\Div{div}

\DeclareMathOperator\rot{curl}

\renewcommand\leq\leqslant
\renewcommand\geq\geqslant
\newlength{\intwidth}

\def\tw{\tilde w}
\begin{document}

\title[Knots and links in steady solutions of the Euler equation]{Knots and links in steady solutions\\ of the Euler equation}

\author{Alberto Enciso}
\address{Instituto de Ciencias Matemáticas, CSIC-UAM-UC3M-UCM, C/ Serrano 123, 28006 Madrid, Spain}
\email{aenciso@icmat.es, dperalta@icmat.es}

\author{Daniel Peralta-Salas}

%
%
\begin{abstract}
  Given any possibly unbounded, locally finite link, we show that there exists a smooth diffeomorphism transforming this link into a set of stream (or vortex) lines of a vector field that solves the steady incompressible Euler equation in~$\RR^3$. Furthermore, the diffeomorphism can be chosen arbitrarily close to the identity in any $C^r$ norm.\\


\end{abstract}
\maketitle

\section{Introduction}
\label{S.intro}

The three-dimensional incompressible Euler equation
\begin{equation*}
  \frac{\pd u}{\pd t}+(u\cdot \nabla)u=-\nabla P\,,\qquad \Div u=0
\end{equation*}
describes the motion of an inviscid incompressible fluid. A steady solution of the Euler equation is a time-independent vector field $u$ that satisfies the above equation for some pressure function $P$. Equivalently, the vector field $u$ satisfies
\begin{equation*}
  u\wedge\rot u=\nabla B\,,
\end{equation*}
where $B:= P+|u|^2/2$ is the Bernoulli function. In fluid mechanics, the trajectories of the velocity field $u$ and of its curl (i.e., the vorticity) are called {\em stream lines} and {\em vortex lines}, respectively.

Since Lord Kelvin, steady solutions of the Euler equation having knotted and linked stream (or vortex) lines have attracted considerable attention, partly because of their connections with turbulence and hydrodynamic instability~\cite{BR96,Kh05}. Because of the connection of the steady Euler equation with variations of the energy functional under volume-preserving diffeomorphisms, the presence of periodic trajectories of prescribed link type has also been considered to derive lower bounds for the energy functional in certain natural classes of vector fields. These bounds can actually be expressed in terms of the helicity and asymptotic linking number of a fixed vector field of the class (cf.\ e.g.~\cite{FH91,Vo03}  and references therein).

Despite the large body of related literature, there are surprisingly few results of existence of steady solutions with linked stream lines, and in fact our knowledge is mostly based on numerical simulations and exact solutions. In the mid eighties, Moffatt~\cite{Mo85} proposed an appealing heuristic approach to the construction of steady solutions with stream lines diffeomorphic to a given link. Unfortunately, this approach requires global well-posedness for a magnetohydrodynamic system of PDEs including a Navier--Stokes equation (which, as is well known, is very far from obvious) and makes strong topological assumptions on the asymptotic behavior of the solutions as $t\to\infty$.

To our best knowledge, there are only two existence results of steady solutions with prescribed topological properties. Firstly, using contact topology, Etnyre and Ghrist~\cite{EG00} showed that, given a certain link, there exists a (possibly incomplete) Riemannian metric in $\RR^3$ adapted to it such that the Euler equation in this metric admits a steady solution having this link as a set of periodic stream lines. Considering the Euler equation for arbitrary Riemannian metrics is key in order to make the problem amenable to a purely topological approach, so this method cannot be used to derive existence results for a fixed (e.g., Euclidean) metric.

The second result, by Laurence and Stredulinsky~\cite{LS00}, utilizes variational techniques to analyze steady solutions of the Euler equation with axial symmetry. This symmetry makes the problem effectively two-dimensional and allows to formulate it in terms of the scalar stream function. In this case, the authors prove the existence of continuous, energy-class solutions in a bounded domain with level sets of a given topological type. This approach cannot be modified to prescribe stream lines because it provides some (weak) control over the poloidal component of the velocity field but does not yield any information on its axial component. Moreover, at most the so-called torus knots could be possibly realized as stream lines of an axisymmetric solution. 

In this paper we aim to fill this gap by proving the existence of steady solutions in $\RR^3$ with stream or vortex lines of prescribed link type. Here by a {\em link}\/ we mean a disjoint union of tame knots, that is, of smoothly embedded circles in $\RR^3$. The steady solutions we construct belong to the class of (strong) {\em Beltrami fields}~\cite[Section 2.3]{MB02}, i.e., they satisfy
\begin{equation*}
  \rot u=\la u
\end{equation*}
with $\la$ being a nonzero constant. Beltrami fields also play a preponderant role in magnetohydrodynamics, where they are called force-free fields. It should be noticed that the stream lines of a Beltrami field are also vortex lines.

The motivation to consider Beltrami fields to produce steady solutions with stream or vortex lines of arbitrary link type comes from Arnold's and Etnyre and Ghrist's studies of solutions to the steady incompressible Euler equation when the Bernoulli function $B$ is analytic and nonconstant~\cite{Ar65,EG99}. Under certain dynamical assumptions, these authors showed that in this case the only knots that could possibly be stream or vortex lines of a steady solution to the Euler equation in a bounded domain are those of Wada type; this result also holds in the whole $\RR^3$ provided that the solution vector field satisfies the estimate $|u(x)|+|\rot u(x)|\leq A+B|x|$ for some constants~$A$ and $B$. That a similar result does not hold for  Beltrami fields was well known through the analysis of particular instances of ABC flows~\cite{He66}. Our main result in this paper is that actually there are no restrictions on the knot types that can be realized as stream lines of a Beltrami field:

\begin{theorem}\label{T.main}
  Let $L\subset\RR^3$ be a possibly unbounded, locally finite link. Then for any real constant $\la\neq0$ one can transform $L$ by a $C^\infty$ diffeomorphism $\Phi$ of $\RR^3$ arbitrarily close to the identity in any $C^r$ norm, so that $\Phi(L)$ is a set of stream lines of a Beltrami field $u$, which satisfies $\rot u=\la u$ in $\RR^3$.
\end{theorem}

The proof of this theorem is given in Section~\ref{S.proof}, the demonstrations of several intermediate results being postponed to Sections~\ref{S.Cauchy}--\ref{S.global}. An informal guide to the proof of the theorem is presented in Section~\ref{S.strategy}. The proof combines techniques from ordinary and partial differential equations with differential topology. While it is not easy to find parts of the proof purely relying on methods from just one of these areas, the basic philosophy is to use differential topology and ordinary differential equations (as opposed to Laurence--Stredulinsky) to gain control on the trajectories of various auxiliary vector fields, and partial differential equations to relate these auxiliary vector fields to the Beltrami equation (as opposed to Etnyre--Ghrist). 

Theorem~\ref{T.main} provides a rigorous proof of Moffatt's heuristic insight~\cite{Mo85} that any knot should be realizable as a stream line (or vortex line) of a steady solution of the Euler equation. It also furnishes a positive answer to Etnyre and Ghrist's question of whether there exists a steady solution in $\RR^3$ having linked stream lines of all isotopy types; details and further applications will be given in Section~\ref{S.examples}.

It should be noticed that any vector field satisfying $\rot u=\la u$ has infinite
energy, since in this case $\De u=-\la^2u$ and the Laplacian does not
have any $L^2$ eigenfunctions in~$\RR^3$. While there are $L^\infty$
solutions of the equation $\rot u=\la u$ in~$\RR^3$, such as the ABC
flows~\cite{MB02}, the solutions given by Theorem~\ref{T.main} are not
granted to be bounded either. Incidentally, the related problem of
whether there are any nonzero $L^\infty$ steady solutions to the three-dimensional Navier--Stokes equation is wide open; the fact that no bounded solutions exist either in two dimensions or in the axisymmetric case has been established recently~\cite{Acta}.


\section{Strategy of proof}
\label{S.strategy}

Leaving technicalities aside for the moment, the strategy of the proof of the main theorem is the following. We take a connected component $L_a$ of the link $L$, which is a smooth knot, and perturb it a little if necessary to make it real analytic. We then embed the perturbed knot into an analytic two-dimensional strip and construct a vector field on this strip having the latter knot as a stable hyperbolic limit cycle. Next we extend this vector field on the strip to a local Beltrami field  $v_a$, defined in a tubular neighborhood of the cycle. By construction, the field $v_a$ has a periodic orbit diffeomorphic to the link component $L_a$, which is granted to be hyperbolic by the choice of the field on the strip and the fact that the local Beltrami field $v_a$ is divergence-free. The above procedure is then repeated for all the components of the link. To conclude, we show that there exists a global Beltrami field $u$ that approximates each local field $v_a$; the field $u$ will then have a stream line diffeomorphic to each link component $L_a$ because of the robustness of the limit cycle of $v_a$.


Three main difficulties arise when trying to carry out this program. In the first place, it seems natural to define the local Beltrami field $v_a$ as the solution to some Cauchy problem for the curl operator, but the fact that the symbol of $\rot$ is an antisymmetric matrix (and thus there are no non-characteristic surfaces of $\rot$) precludes a straightforward application of the Cauchy--Kowalewski theorem. In the second place, the approximation result connecting the global Beltrami field $u$ and the local fields $v_a$ should be sufficiently fine to effectively deal with the noncompactness of the link $L$. Thirdly and related, the vector field on the strip used as ``Cauchy data'' to define the local Beltrami field $v_a$ must be delicately chosen in order to ensure that the limit cycle of $v_a$ diffeomorphic to the link component $L_a$ is ``robust enough'' to be preserved by the latter approximation.

The proof of Theorem~\ref{T.main}, which we present in the following section, is divided into five steps. Let us very briefly comment on the purpose of each step:

\begin{enumerate}
\item First of all, in Step~1 we present an abstract local existence theorem (Theorem~\ref{T.Cauchy}) for the equation $\rot v=\la v$ with data $v|_\Si=w$ on a surface $\Si$ that will be subsequently applied to define the local Beltrami field $v_a$ in terms of a prescribed vector field on a strip at the link component $\tL_a$. To circumvent the lack of ellipticity of $\rot$ (or of its counterpart $\star d$ acting on $1$-forms, with $\star$ the Hodge operator), we resort to an indirect argument with the Dirac-type operator $d+d^*$ acting on the space of differential forms of mixed degrees. Here $d^*$ is the codifferential; the operator $d+d^*$ is then elliptic (thus allowing us to invoke the Cauchy--Kowalewski theorem), but it does not preserve the space of $1$-forms. Ultimately, this reflects itself in the necessity of imposing some technical conditions on the vector field on the strip that will serve as Cauchy data.

\item Appropriate Cauchy data (that is, the vector field on the strip at each link component $L_a$  previously mentioned) are carefully constructed in Step~2 (Lemma~\ref{L.local}). Each field is required to satisfy the technical conditions demanded by the existence theorem proved in Step~1 and, besides, to have a stable hyperbolic limit cycle diffeomorphic to the link component $L_a$. This construction is considerably simplified by the introduction of a local coordinate system adapted to the link component.

\item The choice of Cauchy data on the strip we made in Step~2 ensures that the associated local Beltrami field $v_a$ obtained using the existence theorem of Step~1 also has a periodic orbit diffeomorphic to the link component $L_a$. In Step~3 we show that this periodic orbit is robust under suitably small perturbations of the vector field because our choice of Cauchy data guarantees that there is a stable manifold tangent to the strip and, the field $v_a$ being divergence-free, there must also be an unstable manifold, thereby granting that the periodic orbit is hyperbolic (Theorem~\ref{T.eta}).

\item In Step~4 we prove an abstract result on approximation of local Beltrami fields by global Beltrami fields (Theorem~\ref{T.Carleman}) adapted to the kind of perturbations that were allowed in Step~3. This result relies on a better-than-uniform approximation theorem for the auxiliary scalar  elliptic equation $\De f=-\la^2 f$, which in turn hinges on an iterative scheme that employs the Lax--Malgrange theorem.

\item To conclude, in Step~5 we use the latter approximation theorem to construct a global Beltrami field $u$ that approximates each local Beltrami field $v_a$ in a neighborhood of the link component $L_a$ and show that therefore the field $u$ has a set of periodic trajectories diffeomorphic to the link $L$ on account of the hyperbolic permanence theorem.
\end{enumerate}

\section{Proof of the main theorem}
\label{S.proof}

In this section we shall present the proof of
Theorem~\ref{T.main}, which, as we have just mentioned, is divided into five steps. We will henceforth assume that the parameter $\la$ that appears in the Beltrami equation $\rot u=\la u$ is positive (which can be accomplished by mapping $x\mapsto -x$ if necessary) and that all the diffeomorphisms are of class $C^\infty$.

\subsubsection*{Step 1 (Local existence for the Cauchy problem)}

A natural way of obtaining local solutions of the Beltrami equation with partially prescribed behavior is to consider the Cauchy problem
\begin{equation*}
  \rot v=\la  v\,,\qquad v|_{\Si}=w\,,
\end{equation*}
where $\Si$ is an embedded oriented surface in~$\RR^3$ of class $C^\om$ (where, as customary, $C^\om$ stands for real analytic). Therefore, our goal in this step is to establish the following abstract existence theorem for the above Cauchy problem, which grants the existence of a solution under suitable hypotheses on the vector field  $w$ used as Cauchy data:


\begin{theorem}\label{T.Cauchy}
  Let $\Si$ be an embedded oriented analytic surface in $\RR^3$. Let $w$ be a $C^\om$
  vector field tangent to $\Si$ and let us denote by $\ga$ its associated $1$-form. If the pullback of the $1$-form $\ga$ to the surface $\Si$ is closed (i.e., $dj_\Si^*(\ga)=0$ with $j_\Si:\Si\to\RR^3$ the inclusion map), then the equation $\rot v=\la v$ with Cauchy data $v|_\Si=w$ has a unique solution in a neighborhood of the surface $\Si$. This solution is analytic.
\end{theorem}

The proof of this result is given in Section~\ref{S.Cauchy}. To understand the content of the theorem, it is worth discussing at this point the theorem's hypothesis that the pulled-back $1$-form $j_\Si^*(\ga)$ is closed. To this end, let us take $C^\om$ local coordinates $(\rho,\xi^1,\xi^2)$ in a neighborhood of the surface $\Si$, where $\rho$ denotes the signed distance to this surface~\cite{KP81}. In these coordinates, the Euclidean metric reads
\begin{equation}\label{rhoxi}
  ds^2=d\rho^2+ h_{ij}(\rho,\xi)\,d\xi^i\,d\xi^j\,,
\end{equation}
and obviously the surface $\Si$ is the zero set of the coordinate $\rho$. We will denote by $h^{ij}$ the inverse matrix of $h_{ij}$ and by $|h|:=\det(h_{ij})$ its determinant. Let us write the vector field $v$ in these coordinates as
\[
v=\chi(\rho,\xi)\,\frac\pd{\pd\rho}+ b^i(\rho,\xi)\,\frac\pd{\pd\xi^i}\,;
\]
setting  $a_i:=h_{ij}b^j$, its associated $1$-form can be written as
\[
\be:=\chi(\rho,\xi)\,d\rho+ a_i(\rho,\xi)\,d\xi^i\,.
\]

Let us suppose that the field $v$ satisfies the equation $\rot v=\la v$ and is tangent to the surface $\Si$, as claimed in the theorem. Since the coordinate vector field $\pd/\pd\rho$ is orthogonal to $\pd/\pd\xi^i$ by the coordinate expression of the metric (cf.\ Eq.~\eqref{rhoxi}), the condition that the field $v$ be tangent to the surface is obviously tantamount to demanding that the function $\chi$ appearing in the coordinate expression of $v$ satisfies $\chi(0,\xi)=0$ for all $\xi$.

It is slightly more convenient to attack the equation $ \rot v=\la  v$ using differential forms, in terms of which it can be rewritten as $\star d\be=\la \be$, where $\be$ is the $1$-form dual to the field $v$, $\star$~denotes the Hodge operator and $d$ is the exterior derivative. The action of $\star d$ on the $1$-form $\be$ is given in coordinates by
\begin{multline*}
  \star d\be= |h|^{1/2} h^{1i}h^{2j} \bigg(\frac{\pd a_j}{\pd\xi^i} -\frac{\pd a_i}{\pd\xi^j}\bigg) \,d\rho +|h|^{1/2} h^{2i} \bigg(\frac{\pd \chi}{\pd\xi^i} -\frac{\pd a_i}{\pd\rho}\bigg) \,d\xi^1\\
+|h|^{1/2} h^{1i} \bigg(\frac{\pd a_i}{\pd\rho}-\frac{\pd \chi}{\pd\xi^i}\bigg) \,d\xi^2\,.
\end{multline*}
Accordingly, the equation $\star d\be=\la\be$ and the condition $\chi(0,\xi)=0$ imply that the coefficients $a_i(\rho,\xi)$ of the $1$-form $\be$ satisfy
\[
\frac{\pd a_j}{\pd\xi^i}(0,\xi) -\frac{\pd a_i}{\pd\xi^j}(0,\xi)=0\,.
\]
This means that the pullback  $j_\Si^*(\be)$ to the surface of the $1$-form $\be$ must be closed. Hence the content of Theorem~\ref{T.Cauchy} is that this condition (which we state in terms of the Cauchy data) is not only necessary, as the calculations above show, but also sufficient.

\subsubsection*{Step 2 (Construction of appropriate Cauchy data)}

In view of the local existence theorem proved in Step~1, our goal in this step is to construct vector fields tangent to certain analytic surfaces that satisfy the hypotheses of the aforementioned theorem and, in addition, possess certain convenient dynamical properties on these surfaces that will be discussed later. Since we will need to work in the analytic category, as a preliminary technical result we start by recording that there is a ``small'' diffeomorphism $\tilde\Phi$ of $\RR^3$ which transforms the link $L$ into an analytically embedded submanifold $\tilde\Phi(L)$. More precisely,

\begin{lemma}\label{L.embed}
  Given the locally finite link $L$, we can transform it by a diffeomorphism $\tilde\Phi$ of $\RR^3$ arbitrarily close to the identity in any $C^r$ norm, so that $\tilde\Phi(L)$ is a real analytic link.
\end{lemma}

We omit the proof of this result, which is a straightforward consequence of Thom's isotopy theorem  (cf.\ e.g.~\cite[Theorem 14.1.1]{BCR98}). Replacing the link $L$ by its transformed image $\tilde\Phi(L)$ if necessary, by the above lemma we can (and will) assume that the link $L$ is analytic without loss of generality. We will denote the connected components of the link by $\tL_a$, where the labeling index $a$ ranges over an at most countable set $A$.

In order to apply the Existence Theorem~\ref{T.Cauchy} later on, we need to introduce an appropriate surface $\Si_a$ containing the link component~$L_a$. As the normal bundle of $\tL_a$ is trivial~\cite{Ma59}, a technically convenient way of doing this is via a $C^\om$ trivializing map $\Theta_a$ whose $0$-fiber is $\tL_a$:

\begin{definition}\label{D.strip}
 We say that an embedded analytic surface $\Si_a$ is a {\em strip} at the link component $\tL_a$ if there is a tubular neighborhood $N_a$ of the link component and an associated $C^\om$ trivializing map $\Theta_a:N_a\to\RR^2$ whose $0$-fiber is $\tL_a$ and such that the surface $\Si_a$ is given by $\Theta_a^{-1}((-1,1)×\{0\})$. Obviously the link component $\tL_a$ is contained in the strip $\Si_a$.
\end{definition}

Clearly the surface $\Si_a$ is diffeomorphic to the cylinder $\SS^1×\RR$, so it will be assumed to be oriented. The following lemma provides a way of choosing a vector field $w_a$ tangent to a strip at a link component $\tL_a$ so that, on the one hand, satisfies the hypotheses of the Existence Theorem~\ref{T.Cauchy} that we stated in Step~1 and, on the other hand, its pullback to the surface has the link component  $\tL_a$ as a {\em  stable hyperbolic limit cycle}. This simply means that there exists a neighborhood of the link component in the strip $\Si_a$ whose $\om$-limit set along the local flow of the pulled back vector field is this component and that the associated monodromy matrix does not have any nontrivial eigenvalues of modulus $1$.

\begin{lemma}\label{L.local}
  Given a link component $\tL_a$, one can take a tubular neighborhood $N_a$ of $\tL_a$ and an associated strip $\Si_a$ such that there is an analytic vector field $w_a$ in $N_a$ tangent to the strip and whose pullback to the strip (i.e., $j_a^*(w_a)$ with $j_a:\Si_a\to N_a$ the inclusion map) has the link component $\tL_a$ as a stable hyperbolic limit cycle. Furthermore, one can assume that the pullback to the strip $\Si_a$ of the $1$-form associated to the field $w_a$ is closed and that tubular neighborhoods corresponding to distinct link components have disjoint closures.
\end{lemma}

The proof of Lemma~\ref{L.local} is presented in Section~\ref{S.local}. One should note that the fact that the tubular neighborhoods corresponding to distinct link components have disjoint closures will be used in Step~5 to construct the diffeomorphism $\Phi$ which transforms the link into a set of stream lines.

\subsubsection*{Step 3 (Stability of periodic stream lines)}

Given a link component $\tL_a$, we showed in Step~2 (Lemma~\ref{L.local}) that one can take an appropriate surface $\Si_a$ containing~$\tL_a$, which we called a strip, and a vector field $w_a$ tangent to the strip which has certain convenient dynamical properties and satisfies the requirements of the Existence Theorem~\ref{T.Cauchy} proved in Step~1. Therefore, there is a unique analytic local Beltrami field $v_a$, which satisfies the equation $\rot v_a=\la v_a$ in a small enough tubular neighborhood $N_a$ of the link component, whose restriction to the strip is $v_a|_{\Si_a}=w_a$.

By construction, the link component $\tL_a$ is a periodic orbit of the local Beltrami field $v_a$. Our main result in this step is the following theorem, which asserts that this periodic orbit is robust under suitably small perturbations.  The basic idea is that, as our choice of the vector field $w_a$ on the strip guarantees that the cycle~$\tL_a$ of the field $v_a$ has a stable manifold (which is tangent to the strip), the fact that the field $v_a$ preserves volume implies that the cycle also has an unstable manifold, thus ensuring its hyperbolicity.  The proof of this result is given in Section~\ref{S.stability}.

\begin{theorem}\label{T.eta}
  Consider a tubular neighborhood $N_a$ of the link component $\tL_a$, and an associated strip $\Si_a$. Let the field $v_a$ be the unique solution to the equation $\rot v_a=\la v_a$ in this neighborhood taking the value $v_a|_{\Si_a}=w_a$, where $w_a$ is the vector field constructed in Lemma~\ref{L.local}. Then if $\tv$ is any vector field $C^r$-close to the field $v_a$ in the neighborhood $N_a$ with $r\geq1$, there is a diffeomorphism $\Phi_a$ of $\RR^3$ $C^r$-close to the identity transforming the link component~$L_a$ (which is a periodic orbit of the field $v_a$) into a periodic orbit $\Phi_a(\tL_a)$ of the perturbed vector field $\tv$. Furthermore, one can assume that the diffeomorphism $\Phi_a$ coincides with the identity outside the open set $N_a$.
\end{theorem}

\subsubsection*{Step 4 (Better-than-uniform approximation by global solutions)}

Our goal in this section is to present an abstract theorem that permits to approximate a local Beltrami field by a global Beltrami field. The paradigmatic result on the approximation of local solutions of an analytic differential equation by global solutions is the celebrated Lax--Malgrange theorem~\cite{La56,Ma56}, which applies to linear elliptic differential operators and yields uniform approximation on compact sets. The hypotheses of compactness and, most importantly, ellipticity are certainly not satisfied by the Beltrami equation, so our analysis requires some further elaboration. Moreover, better-than-uniform approximation is necessary in order to exploit the stability results we carefully established in Step~3.

Better-than-uniform approximation refers to the fact that we do not only wish to prove that the $C^r$ norm of the difference of two vector fields is small (say, smaller than a certain number $\ep$) but, rather, that the $C^r$-approximation can be actually improved as one moves away from a fixed point (e.g., the origin). This is usually expressed using an ``error function'' $\ep(x)$ and pointwise bounds, as we do in the statement of this theorem. Its proof will be provided in Section~\ref{S.global}. 

\begin{theorem}\label{T.Carleman}
  Let the closed set $S$ be a locally finite union of pairwise disjoint compact subsets of~$\RR^3$ whose complements are connected, and let $v$ be a local Beltrami field that satisfies $\rot v=\la v$ in $S$. Then one can find a $C^r$ better-than-uniform approximation of the field $v$ by a global Beltrami field $\tv$, i.e., for any integer $r$ and any positive continuous function $\ep(x)$ from $S$ to $(0,\infty)$ there is a field $\tv$ which satisfies the equation $\rot \tv=\la\tv$ in $\RR^3$ and such that its difference with the field $v$ admits the pointwise $C^r$ bound
\begin{equation*}
\sum_{|\al|\leq r}\big|D^\al v(x)-D^\al\tv(x)\big|<\ep(x)
\end{equation*}
in the set $S$.
\end{theorem}

Regarding the hypotheses on the set $S$ in the statement of the theorem, it should be mentioned that the condition that the complement of each compact subset be connected is necessary in order to apply the Lax--Malgrange theorem for an auxiliary elliptic equation, while the fact that the set $S$ is the locally finite union of disjoint compact sets is used to attack the full problem using an iterative scheme. The assumption that the equation $\rot v=\la v$ holds in the {\em closed}\/ set $S$ (which means that it holds in an open neighborhood of $S$ and is also a hypothesis of the Lax--Malgrange theorem) is required to ensure that the fields do not have a pathological behavior at the boundary.

\subsubsection*{Step 5 (Construction of the global Beltrami field)}

Finally, in this step we will obtain the desired global Beltrami field $u$ having a set of periodic stream lines diffeomorphic to the link $L$ using the local Beltrami fields $v_a$ constructed in Step~3 and the approximation theorem established in Step~4. By taking smaller neighborhoods if necessary, we will assume that the latter fields satisfy the equation $\rot v_a=\la v_a$ in the closure $\overline{N_a}$ of a tubular neighborhood of the link component $L_a$. In order to apply the Approximation Theorem~\ref{T.Carleman}, we define a vector field $v$ in the union $S:=\bigcup_{a\in A} \overline{N_a}$ of all these sets by letting it be equal to the local Beltrami field $v_a$ in each set $\overline{N_a}$. Here we are using that these sets $\overline{N_a}$ can be assumed disjoint for distinct link components, as we saw in Lemma~\ref{L.local}.

Since the field $v$ thus defined satisfies the Beltrami equation $\rot v=\la v$ in the closed set $S$ and $S$ verifies the hypotheses of the Approximation Theorem~\ref{T.Carleman}, we infer that there is a global Beltrami field $u$ that approximates the field $v$ in the $C^r$ better-than-uniform sense. That is, given any continuous function $\ep(x)$ mapping the set $S$ into the positive reals $(0,\infty)$ and an integer $r$ (the order of approximation), one can choose the global Beltrami field $u$ so that in the set $S$ we have the pointwise $C^r$ bound
\[
\sum_{|\al|\leq r}\big|D^\al v(x)-D^\al u(x)\big|<\ep(x)\,.
\]

We shall next show that the global Beltrami field $u$ has a set of stream lines diffeomorphic to the link $L$. This will hinge on the robustness of the periodic orbits corresponding to each link component $L_a$ of the local Beltrami fields $v_a$, proved in Theorem~\ref{T.eta} of Step~3 for suitably $C^r$-small perturbations with $r\geq1$. More precisely, in the present situation this theorem asserts that for any $\de>0$ and any link component~$L_a$ one can take a positive constant $\ep_a$ (depending on $\de$ and on the component) such that, if the vector field $u$ is $\ep_a$-close in the $C^r$ norm (with $r\geq 1$) to the local Beltrami field $v_a$ in the set $N_a$, then there is a diffeomorphism $\Phi_a$ of $\RR^3$ transforming the link component $L_a$ into a periodic stream line $\Phi_a(L_a)$ of the global Beltrami field $u$ and which lies in a $\de$-neighborhood of the identity (that is, $\|\Phi_a-\id\|_{C^r(\RR^3)}<\de$). Since the field $v$ coincides with the local Beltrami field $v_a$ in the set $\overline{N_a}$ (for each labeling index $a$), one can certainly ensure that the global Beltrami field $u$ is sufficiently close to the field $v_a$ in each neighborhood $N_a$ by using the aforementioned $C^r$ better-than-uniform approximation result. (More precisely, this amounts to taking an order of approximation $r\geq1$ and an ``error function'' $\ep(x)$ which is bounded from above by the previously introduced constant $\ep_a$ in each set $\overline{N_a}$.) It should be noticed that better-than-uniform approximation is essential for the argument, as the constants $\ep_a$ might not be uniformly bounded away from zero (with respect to the index $a$).

We saw in the Stability Theorem~\ref{T.eta} that the above diffeomorphism $\Phi_a$ was only different from the identity in the open neighborhood $N_a$ of the link component $L_a$, so these diffeomorphisms can be trivially ``glued together'', defining a diffeomorphism $\Phi$ of $\RR^3$ as
\[
\Phi(x):=\begin{cases}  \Phi_a(x) &\text{if } x\in N_a\,,\\
   x &\text{if }x\not\in \bigcup_{a\in A}N_a\,.
\end{cases}
\]
This is the diffeomorphism whose existence was claimed in the Main Theorem~\ref{T.main}, for it transforms the link $L$ into a set of periodic stream lines $\Phi(L)$ of the global Beltrami field $u$ (which satisfies the equation $\rot u=\la u$ in $\RR^3$) and, by construction, the $C^r$-distance between this diffeomorphism and the identity is at most $\de$.

\section{Proof  of Theorem~\ref{T.Cauchy}}
\label{S.Cauchy}

In this section we will prove Theorem~\ref{T.Cauchy} of Step~1, which ensures existence and uniqueness of solutions to the equation $\rot v=\la v$ with Cauchy data $v|_\Si=w$ on the surface $\Si$ (or, in terms of the associated $1$-forms, $\star d\be=\la\be$ and $\be|_\Si=\ga$) under appropriate hypotheses on the vector field $w$. We will need to use differential forms of mixed degrees, so for any domain $V\subset\RR^3$ we use the notation
\[
\Om^\bullet(V):=\bigoplus_{p=0}^3\Om^p(V)
\]
for the graded algebra of $C^\infty$ differential forms in this domain. The elements of $\Om^\bullet(V)$ will be denoted as
\[
\psi=\psi^0\oplus \psi^1\oplus \psi^2\oplus \psi^3\,,
\]
and we henceforth identify a $p$-form with its natural inclusion in $\Om^\bullet(V)$.
The parity operator $Q:\Om^\bullet(V)\to\Om^\bullet(V)$ is defined by
\[
Q\psi:=\psi^0\oplus -\psi^1\oplus \psi^2\oplus -\psi^3\,.
\]

Clearly the differential operator $\rot$ (or its counterpart $\star d$, which maps the space of $1$-forms $\Om^1(\RR^3)$ into itself) is not immediately amenable to an approach based on the Cauchy--Kowalewski theorem because it has a degenerate symbol. Hence, we find it convenient to consider the Dirac-type operator $d+d^*$ acting on $\Om^\bullet(\RR^3)$, where $d^*$ denotes the codifferential. Notice that the operator $d+d^*$ is elliptic, as its iterated square equals the Laplacian $-\De$, but the price to pay for this is that, unlike the ``curl'' $\star d$, $d+d^*$ does not generally map $1$-forms into $1$-forms.

We can invoke the Cauchy--Kowalewski theorem to derive that, for some neighborhood $V$ of the surface $\Si$, there exists a unique $C^\om$ solution $\psi\in\Om^\bullet(V)$ of the following auxiliary Cauchy problem:
\begin{equation*}
  (d+d^*)\psi=\la\star\psi\,,\qquad \psi|_{\Si}=\ga\,,
\end{equation*}
where $\ga$ is the $1$-form associated with the vector field $w$ appearing in the statement of Theorem~\ref{T.Cauchy} as Cauchy data. 

A crucial observation is the following technical lemma, which asserts that the hypotheses made on the Cauchy data imply that the unique solution of the above problem is co-closed (i.e., $d^*\psi=0$). This will be used to relate solutions of the above auxiliary Cauchy problem to solutions of the equation $\rot v=\la v$, $v|_\Si=w$:

\begin{lemma}\label{L.div}
  Under the hypotheses on the $1$-form $\ga$ imposed in Theorem~\ref{T.Cauchy} (namely, that its pullback to the surface is closed and that its dual vector field is tangent to the surface), the solution of the Cauchy problem $(d+d^*)\psi=\la\star\psi$ with $\psi|_{\Si}=\ga$ is co-closed, i.e.,  $d^*\psi=0$.
\end{lemma}
\begin{proof}
  By the identity $\star d\star =-Q d^* $, taking the action of $\star d$ on the equation $(d+d^*)\psi=\la \star\psi$ we readily obtain that the form $d^*\psi$ satisfies the PDE
\[
(d+d^*)d^*\psi=-\la\star Q\,d^*\psi\,.
\]
The Cauchy--Kowalewski theorem will then yield that the codifferential of the form $\psi$ is identically zero in its domain $V$ provided that it vanishes on the surface $\Si$.

  In order to show that the codifferential $d^*\psi$ is zero on $\Si$, let us use the local coordinates $(\rho,\xi^1,\xi^2)$ introduced in Step~1 of Section~\ref{S.proof} and denote by $\bar d$ the exterior derivative in the variables $(\xi^1,\xi^2)$, so that the hypothesis that the $1$-form $j_\Si^*(\ga)$ is closed is equivalent to the condition $\bar d\ga|_\Si=0$.  Let us decompose the form $\psi$ as
  \[
\psi=:\Psi+d\rho\wedge \tPsi
  \]
  for some forms $\Psi,\tPsi\in\Om^\bullet(V)$ without $d\rho$ (by which we mean that the interior product of the coordinate vector field $\pd/\pd\rho$ with them vanishes identically). As $\psi|_\Si=\ga$ is a $1$-form without $d\rho$ and one has $\bar d\ga|_\Si=0$ by hypothesis, it is easy to check that, on the surface, the star of the form $\psi$ can be written as
  \[
\star\psi|_\Si=\star\ga|_\Si=:d\rho\wedge\tilde\ga
  \]
  for some $1$-form $\tilde\ga$, while its differential reads as
  \[
d\psi|_\Si=\bigg[\bar d\Psi+d\rho\wedge\bigg(\frac{\pd\Psi}{\pd\rho}-\bar d\tPsi\bigg)\bigg]\bigg|_\Si =d\rho\wedge\frac{\pd\Psi}{\pd\rho}\,.
\]
The codifferential of the form $\psi$ on the surface can then be computed using the equation $(d+d^*)\psi=\la\star\psi$, finding that
\begin{equation}\label{referee1}
d^*\psi|_\Si=\big(\la\star\!\psi-d\psi\big)\big|_\Si= d\rho\wedge\bigg(\la\tilde\ga-\frac{\pd\Psi}{\pd\rho}\bigg)\,.
\end{equation}

The degree-$0$ component of the right hand side of the above equation is necessarily $0$, which implies that the codifferential of the degree-1 component of $\psi$ vanishes on the surface, i.e., $d^*(\psi^1)|_\Si=0$. For degrees $p=2,3$, the fact that the form $\psi^p$ is zero on the surface $\Si$ implies that
\[
d^*(\psi^p)|_\Si=(-1)^p\big(\star d\star \psi^p\big)\big|_\Si= (-1)^p \star \bigg(d\rho\wedge \frac{\pd\star\! \psi^p}{\pd \rho}\bigg)\,,
\]
which is a $(p-1)$-form without $d\rho$. Comparison with the degree-$(p-1)$ component of $d^*\psi$, computed in Eq.~\eqref{referee1}, then shows that $d^*(\psi^p)|_\Si=0$ for $p=2,3$ and the lemma follows.
\end{proof}

As the form of mixed degree $\psi$ solving $(d+d^*)\psi=\la\star\psi$ with $\psi|_\Si=\ga$ is co-closed by Lemma~\ref{L.div} above, from the identity $\star( \star \psi)=\psi$ we deduce that the form $\psi$ also satisfies the equation $\star d\psi=\la\psi$. Since $\star d$ maps $1$-forms into $1$-forms, the degree-1 component $\be:=\psi^1$ is then the unique solution of the Cauchy problem
\[
\star d\be=\la\be\,,\qquad \be|_\Si=\ga
\]
in a neighborhood of the surface, and hence Theorem~\ref{T.Cauchy} follows by letting $v$ be the vector field associated to the $1$-form $\be$ via the Euclidean metric.

\section{Proof of Lemma~\ref{L.local}}
\label{S.local}

In this section we construct a vector field $w_a$, tangent to a strip $\Si_a$ contained in a tubular neighborhood $N_a$ of the link component $L_a$, which satisfies the hypotheses of the Existence Theorem~\ref{T.Cauchy} and whose pullback to the surface has the link component as a hyperbolic limit cycle. As we will see, the $1$-form $\ga_a$ dual to the desired field $w_a$ is simply given by $d\theta-z\,dz$ in a natural coordinate system; hence, we will start by carefully defining these coordinates on the neighborhood $N_a$. The coordinates we shall use are $(\rho,z,\theta)$, where $\rho$ is the signed distance to the surface $\Si_a$ and, roughly speaking, $z$ and $\theta$ are suitable extensions of the signed distance to the link component $L_a$ as measured on the surface and of the arc-length parameter of $L_a$, respectively.

To construct the coordinates, we begin with an arc-length parametrization of the link component $\tL_a$, which is a periodic function $\theta$ mapping the component to the reals modulo the length $|\tL_a|$ of the link component (i.e., $\RR/|L_a|\ZZ$). By taking the strip small enough, we can safely assume that the signed distance function to the link component $\tL_a$, as measured along the surface $\Si_a$ with respect to the metric $h $ inherited from its embedding in $\RR^3$, is analytic. Let us denote this function by $z :\Si_a\to\RR$. The gradient operator  associated to the metric $h $ on the strip will be denoted by~$\bar\nabla$.

We can extend the parametrization $\theta $ of the link component to a $C^\om$ function on the surface, which we still call $\theta$ with a slight abuse of notation. Calling $\hat\phi_t $ the local flow of the gradient field $\bar\nabla z $ in the surface $\Si_a$, this extension is defined by setting
\[
\theta (\hat\phi_t x):=\theta (x)
\]
for all points $x$ in the link component $\tL_a$ and all $t$ for which $\hat\phi_t x$ is a well-defined point in the surface $\Si_a$. By periodicity, this extension takes values in the reals modulo the length $|L_a|$. Obviously $\bar\nabla\theta \neq0$ in the surface $\Si_a$, and for any point $y\in\Si_a$ there is a unique pair  $(x,t)\in\tL_a×\RR$ such that $y=\hat\phi_t x$ (in fact, here $t$ is the signed distance $z (y)$ because $h (\bar\nabla z ,\bar\nabla z )=1$). Since the gradients of the functions $\theta$ and $z$ are orthogonal by construction, i.e.,
\begin{equation}\label{ha}
h \big(\bar\nabla \theta ,\bar\nabla z \big)=0\,,
\end{equation}
it is apparent that $(z ,\theta )$ is a coordinate system on the surface $\Si_a$.

To construct the adapted coordinate system in the tubular neighborhood $N_a $, let us begin by considering the signed distance function to the surface $\Si_a$ in $N_a $, which we denote by $\rho$. This function is analytic if the neighborhood $N_a$ is small enough~\cite{KP81}. We now extend the coordinate $z$ (resp.\ $\theta$)   on the surface to a $C^\om$ function mapping the neighborhood $N_a$ into the reals (resp., by periodicity, the reals modulo the length $|\tL_a|$); for simplicity, we still denote the extended functions by $z$ and $\theta$. This extension is performed using the local flow $\phi_t $ of the gradient field $\nabla\rho $ as
\[
z (\phi_t x):=z (x),\qquad \theta (\phi_t x):=\theta (x)\,,
\]
for all points $x\in\Si_a$ and values $t$ such that $\phi_t x\in N_a $ is defined. Arguing as above, it can be readily established that $(\rho ,z ,\theta )$ is an analytic coordinate system in the open set $N_a $ verifying the conditions
\begin{equation}\label{orth}
  \lan\nabla\rho ,\nabla\theta \ran= \lan\nabla\rho ,\nabla z \ran=0\,,\qquad |\nabla\rho |=1\,.
\end{equation}
It should be noticed that the surface $\Si_a$ is given by the zero set $\{\rho =0\}$ and the link component $\tL_a$ is $\{\rho =z =0\}$, which implies the relations
\begin{equation}\label{orth2}
\lan\nabla z ,\nabla\theta \ran\big|_{\Si_a}=0\,,\qquad |\nabla\theta |\big|_{\tL_a}=1
\end{equation}
as a consequence of the orthogonality condition~\eqref{ha} and the fact that $\theta|_{\tL_a}$ is an arc-length parametrization of the link component $L_a$.

As we mentioned, in this coordinate system we can define the $1$-form associated to the desired field $w_a$ as
\[
\ga_a :=d\theta -z \,d z\,.
\]
The pullback to the strip of this $1$-form is obviously closed.  The associated vector field, $w_a :=\nabla\theta -z \,\nabla z $, is clearly tangent to the strip $\Si_a$ because it is orthogonal to $\nabla\rho $ by Eq.~\eqref{orth}. Moreover, by the relations~\eqref{orth2} it follows that the field $w_a $ is nonvanishing and orthogonal to $\nabla z $ on the link component $\tL_a$, which implies that this component is a periodic orbit of the field $w_a$. 

That the link component $\tL_a$ is an asymptotically stable periodic orbit of the pullback $j_a^*(w_a)$ of the field $w_a$ to the surface will stem from the existence of an appropriate Lyapunov function. This Lyapunov function is simply $z^2$, since it is nonnegative, vanishes exactly on the cycle and its scalar product with the field $w_a$,
\[
\lan w_a ,\nabla (z ^2)\ran|_{\Si_a}=2\big(z \lan\nabla z ,\nabla\theta \ran -z ^2|\nabla z |^2\big)\big|_{\Si_a}=-2z ^2|\nabla z |^2|_{\Si_a}\,,
\]
is zero on the cycle $\tL_a=\{z =0\}\cap\Si_a$ and strictly negative in its complement~$\Si_a\minus\tL_a$.

To prove the hyperbolicity of the cycle $\tL_a$, we start by observing that the field $w_a$ on the surface has the form
\begin{equation}\label{referee3}
j_a^*(w_a )=g(z,\theta) \,\frac\pd{\pd\theta }-z \,\frac{\pd}{\pd z }
\end{equation}
on account of the relations~\eqref{orth} and~\eqref{orth2}, where the function $g(z,\theta)$ is defined in terms of the gradient of $\theta$ and the metric $h$ on the surface as $g :=h (\bar\nabla\theta ,\bar\nabla\theta )$. Hence, the integral curve $\Ga (t)$ of the field $j_a^*(w_a )$ corresponding to the link component $\tL_a$ is parametrized by $z (t)=0$ and $\theta (t)=t$ modulo the length $|\tL_a|$. The monodromy matrix along this orbit has only one nontrivial eigenvalue $\mu $, which by Liouville's formula (cf.\ e.g.~\cite[Theorem IV.1.2]{Ha82}) can be expressed as
\begin{align*}
  \mu &=\exp\bigg[\int_0^{|\tL_a|} \bigg(\frac{\pd g }{\pd\theta }\circ\Ga (t)-1\bigg)\,dt\Bigg] =  \e^{-|\tL_a|}<1
\end{align*}
using that the function $g \circ \Ga $ is $|\tL_a|$-periodic. Hyperbolicity then follows.

\section{Proof of Theorem~\ref{T.eta}}
\label{S.stability}

We shall prove the Stability Theorem~\ref{T.eta}, which ensures that the link component~$\tL_a$ is a robust periodic stream line of the local field $v_a$, which solves the equation $\rot v_a=\la v_a$ with Cauchy data $v_a|_{\Si_a}=w_a$ chosen as in Step~2. To do this, we will use the same coordinate system $(\rho ,z ,\theta )$ we introduced in the proof of Lemma~\ref{L.local}, defined in a tubular neighborhood $N_a$ of the cycle.

From the construction of the vector field $w_a$ used as Cauchy data (cf.\ Lemma~\ref{L.local}), it stems that the local Beltrami field $v_a$ can be written as
\[
v_a=\big(g (\theta )+\cO(z )+\cO(\rho )\big)\,\frac\pd{\pd\theta } - \big( z \,f (\theta )+\cO(\rho )+\cO(z ^2)\big)\,\frac\pd{\pd z }+ \cO(\rho )\,\frac\pd{\pd \rho }\,.
\]
In fact, from the expression for the pulled back field $j_a^*(w_a)$ given in Eq.~\eqref{referee3} it follows that we can set $f(\theta)= 1$ above. As the field $v_a$ is analytic, the small terms behave like symbols in the sense that, e.g.,
\[
\bigg|\frac{\pd^{n+m+l}}{\pd\rho ^n\pd z ^m\pd \theta ^l}\cO(\rho ^s)\bigg|\leq F_{mnl}(z ,\theta )\,|\rho |^{s-n}
\]
for some smooth function $F_{mnl}$. The function  $g(\theta) $ entering the above expression for the field $v_a$ is nonvanishing and $|\tL_a|$-periodic, $|\tL_a|$ being the length of the link component, which means that we can rescale the coordinate $\theta $ to have $g(\theta) =1$. The integral curve $\Ga (t)$ of the field $v_a$ corresponding to the link component $\tL_a$, which will have certain period $T$ after the rescaling, is given in these coordinates by $\rho (t)=0$, $z (t)=0$ and $\theta (t)=t$ modulo $T$.

A key intermediate result is that the cycle $\tL_a$ is hyperbolic. The proof is based on a monodromy matrix computation:

\begin{lemma}\label{L.local2}
  The link component $\tL_a$ is a hyperbolic periodic orbit of the local Beltrami field $v_a$.
\end{lemma}
\begin{proof}
The variational equation associated with the integral curve $\Ga (t)$ corresponding to the link component can be written as
\begin{equation*}
\frac{dy}{dt}=Dv_a(\Ga (t))\,y(t)\,,
\end{equation*}
where $y(t)$ is a three-component vector and
\[
Dv_a(\Ga (t))=\left(\begin{matrix}
  F_1(t) & 0 &0\\
  F_2(t) & -1 &0\\
  F_3(t) & F_4(t) &0
\end{matrix}\right)
\]
is the Jacobian matrix of the field $v_a$ in the coordinates $(\rho ,z ,\theta )$ evaluated at the integral curve $\Ga (t)$. Here $F_j(t)$ ($1\leq j\leq 4$) are some analytic $T$-periodic functions whose explicit expression we will not need.

The solution of the variational equation above is given by $y(t)=M(t) \,y(0)$, where
\[
M(t):=\left(\begin{matrix}
  G_1(t) & 0 &0\\
  G_2(t) & \e^{-t} &0\\
  G_3(t) & G_4(t) &1
\end{matrix}\right)
\]
is the fundamental matrix and $G_j(t)$ are analytic functions. By definition, the monodromy matrix is $M(T)$, with $T$ being the period, so its nontrivial eigenvalues are $\mu _1:=G_1(T)$ and $\mu _2:= \e^{-T}$.

The field $v_a$ is divergence-free because it satisfies the equation $\rot v_a=\la v_a$, so a direct application of the Liouville formula shows that the determinant of the monodromy matrix is given by
\[
\mu _1\mu _2=\exp\bigg(\int_0^{T}(\Div v_a)\circ\Ga (t)\,dt\bigg)=1\,.
\]
As the eigenvalue $\mu_2=\e^{-T}$ is smaller than $1$, we deduce that the other eigenvalue $\mu _1$ is bigger than $1$, thus establishing the hyperbolicity of the cycle.
\end{proof}

Let us now conclude the proof of the Stability Theorem~\ref{T.eta}. Since the link component $\tL_a$ is a hyperbolic periodic orbit of the field $v_a$, a straightforward application of the hyperbolic permanence theorem~\cite[Theorem 4.1]{HPS77} shows that any vector field~$\tv$ sufficiently $C^r$-close to the field $v_a$ in the neighborhood $N_a$ has a periodic orbit which is a $C^r$-small deformation of the component $\tL_a$, provided $r\geq1$. More precisely, for any integer $r\geq1$ and any $\de>0$ there exists some $\ep>0$ such that any vector field $\tv$ with $\|v_a-\tv\|_{C^r(N_a)}<\ep$ has a periodic orbit $\La$ in the neighborhood $N_a$, and moreover the cycles $\tL_a$ and $\La$ admit smooth trivializing maps $\Theta_a$ and $\widetilde\Theta$ from $N_a$ to $\RR^2$ such that the orbit $\tL_a$ (resp.\ $\La$) is the zero fiber of the map $\Theta_a$ (resp.\ of $\widetilde\Theta$) and the difference of the latter maps is $C^r$-bounded as $\|\Theta_a-\widetilde\Theta\|_{C^r(N_a)}<\de$. The existence of the desired diffeomorphism $\Phi_a$ of $\RR^3$ then follows from Thom's isotopy theorem~\cite[Theorem 14.1.1]{BCR98}.

\section{Proof of Theorem~\ref{T.Carleman}}
\label{S.global}

In this section we shall prove Theorem~\ref{T.Carleman}, which guarantees that any local Beltrami field $v$ defined in a closed set $S$ satisfying certain technical conditions can be approximated in this set by a global Beltrami field $\tv$ in the $C^r$ better-than-uniform sense. Let us denote by $S_b$ the connected components of the closed set $S$, with $b$ ranging over an at most countable set $B$. Since the set $S$ is the locally finite union of $S_b$, it is clear that there exists an exhaustion $\emptyset=:K_0 \subset K_1\subset K_2\subset\cdots$ by compact sets of $\RR^3$ such that:
\begin{enumerate}
\item The union of the interiors of the sets $K_n$, with $n\in\NN$, is $\RR^3$.
\item For each $n$, the complements of the sets $K_n$ and of $S\cup K_n$ are connected.
  \item If the set $K_n$ meets a component $S_b$ of $S$, then $S_b$ is contained in the interior of $K_{n+1}$.
\end{enumerate}

A key technical lemma is the following better-than-uniform approximation result for the auxiliary scalar elliptic equation $(\De+\la^2)f=0$, which is modeled upon a theorem of Bagby and Gauthier~\cite{BG88}. The proof is based on an indirect induction argument that uses the Lax--Malgrange theorem and the properties of the exhaustion we have just introduced.

\begin{lemma}\label{L.Carleman}
  Let the closed set $S$ be a locally finite union of pairwise disjoint compact subsets of~$\RR^3$ whose complements are connected. If the scalar function $f$ verifies the equation $(\De+\la^2)f=0$ in the set $S$, then one can find a $C^s$ better-than-uniform approximation of the function $f$ by a global solution $g$ of the aforementioned equation. That is, for any integer $s$ and any positive continuous function $\tilde\ep(x)$ from the set $S$ to $(0,\infty)$ there is a function $g$ satisfying the equation $(\De+\la^2)g=0$ in $\RR^3$ and such that its difference with the function $f$ has the pointwise $C^s$-bound
\begin{equation*}
\sum_{|\al|\leq s}\big|D^\al f(x)-D^\al g(x)\big|<\tilde\ep(x)
\end{equation*}
in the set $S$.
\end{lemma}
\begin{proof}
The proof relies on an induction argument that is conveniently presented in terms of a  sequence of positive numbers $(\ep_n)_{n=1}^\infty$ related to the error function $\tilde\ep(x)$ and to the above exhaustion by compact sets $K_n$ through the conditions
  \begin{equation}\label{epn}
\ep_m<\frac1{6\ms\si}\min_{x\in K_{m+1}}\tilde\ep(x)\quad\text{and}\quad \sum_{n=m+1}^\infty\ep_n<\ep_m\,,
\end{equation}
which are required to hold for all $m\geq1$. For later convenience, we set $\ep_0:=0$ and we are dividing by the number $\si:=\card\{\al:|\al|\leq s\}$ of multiindices $\al$ with $|\al|$ smaller than or equal to the order of approximation $s$.

The induction hypothesis is that there are functions $g_n:\RR^3\to\RR$ satisfying the equation $(\De+\la^2)g_n=0$ in $\RR^3$ and such that, for any integer $p\geq1$, the following $C^s$ estimates hold:
\begin{subequations}\label{induction}
  \begin{align}
    \sup_{S\cap (K_{p+1}\minus K_p)} \bigg|D^\al\bigg( f-\sum_{n=1}^pg_n\bigg)\bigg|&<\ep_p\,,\label{ind1}\\
    \sup_{S\cap (K_{p}\minus K_{p-1})} \bigg|D^\al\bigg( f-\sum_{n=1}^pg_n\bigg)\bigg|&<\ep_p+2\ep_{p-1}\,,\label{ind2}\\
    \sup_{K_{p-1}}\big| D^\al g_p\big|&<\ep_p+\ep_{p-1}\,.\label{ind3}
  \end{align}
\end{subequations}
Here and throughout the proof of this lemma, all the multiindices will be assumed to range over the set $|\al|\leq s$ without further mention.

Let us start by noticing that, by the Lax--Malgrange theorem~\cite{La56,Ma56}, there exists a function $g_1:\RR^3\to\RR$ satisfying the equation $(\De+\la^2)g_1=0$ in $\RR^3$ and the $C^s$ estimate $| D^\al(f-g_1)|<\ep_1$ in the set $S\cap K_2$. Since we chose $K_0=\emptyset$ and $\ep_0=0$, it is a trivial matter that the induction hypotheses~\eqref{induction} hold for $p=1$. We shall hence assume that the induction hypotheses hold for all $1\leq p\leq m$ and use this assumption to prove that they also hold for $p=m+1$.

To this end, let us construct a function $f_m$ on the set $S\cup K_m$ by setting $f_m:=0$ in the set $K_m$ and defining $f_m$ on each component $S_b$ of the set $S$ as
\[
f_m|_{S_b}:=\begin{cases}
  f-\sum\limits_{n=1}^mg_n &\text{if }S_b\cap (K_{m+2}\minus\stackrel{\circ}K_{m+1})\neq\emptyset\,,\\
  0 &\text{if }S_b\cap (K_{m+2}\minus\stackrel{\circ}K_{m+1})=\emptyset\,.
\end{cases}
\]
Here $\stackrel{\circ}K_n$ stands for the interior of the set $K_n$. The definition of the exhaustion and the first induction hypothesis~\eqref{ind1} guarantee that the function $f_m$ satisfies the equation $(\De+\la^2)f_m=0$ in its domain and that one has the $C^s$ estimate
\begin{equation}\label{fm}
  \sup_{K_m\cup(S\cap K_{m+1})} \big|D^\al f_m\big| \leq \sup_{S\cap( K_{m+1}\minus K_m)} \bigg|D^\al\bigg( f-\sum_{n=1}^mg_n\bigg)\bigg|<\ep_m\,.
\end{equation}
A further application of the Lax--Malgrange theorem allows us to take a function $g_{m+1}$ which satisfies the equation $(\De+\la^2)g_{m+1}=0$ in $\RR^3$ and which is close to the above function $f_m$ in the sense that
\begin{equation}\label{gm1}
  \sup_{K_{m+2}\cap (S\cup K_m)} \big| D^\al(f_m-g_{m+1})\big|<\ep_{m+1}\,.
\end{equation}

Eq.~\eqref{gm1} above and the way we have defined the function $f_m$ in the set $K_{m+2}\minus K_{m+1}$ ensure that the first induction hypothesis~\eqref{ind1} also holds for $p=m+1$. Moreover, from the relations~\eqref{ind1}, \eqref{fm} and~\eqref{gm1} one finds the following pointwise estimate in the set $S\cap (K_{m+1}\minus K_m)$:
\begin{align*}
  \bigg|D^\al\bigg( f-\sum_{n=1}^{m+1}g_n\bigg)\bigg|&\leq \bigg|D^\al\bigg( f-\sum_{n=1}^{m}g_n\bigg)\bigg| + \big|D^\al g_{m+1}\big|\\
  &<\ep_m+ \big| D^\al(f_m-g_{m+1})\big|+ \big|D^\al f_m\big|<\ep_{m+1}+2\ep_m\,.
\end{align*}
This proves the second induction hypothesis~\eqref{ind2} for $p=m+1$. Furthermore,
\[
\sup_{K_m}\big|D^\al g_{m+1}\big|\leq \sup_{K_m}\big|D^\al (f_m-g_{m+1})\big|+ \sup_{K_m}\big|D^\al f_m\big|<\ep_{m+1}+\ep_m
\]
by the relations~\eqref{fm} and~\eqref{gm1}, so the third induction hypothesis~\eqref{ind3} also holds for $p=m+1$. This completes the induction argument.

The desired global solution $g$ can now be defined as
\[
g:=\sum_{n=1}^\infty g_n\,,
\]
with this sum converging $C^s$-uniformly by the definition of the constants $\ep_n$ (see conditions~\eqref{epn}) and the third induction hypothesis~\eqref{ind3}. As the functions $g_n$ verify the equation, it is easily checked that the function $g$ also satisfies the equation $(\De+\la^2)g=0$ in $\RR^3$, which in turn ensures that $g$ is analytic by elliptic regularity. In addition to this, from the definition of the constants~\eqref{epn} and the induction hypotheses~\eqref{induction} it follows that, for any integer $m$, in the set $S\cap (K_{m+1}\minus K_m)$ we have the pointwise $C^s$ estimate
\begin{align*}
 \sum_{|\al|\leq s} \big|D^\al(f-g)\big|  &\leq \sum_{|\al|\leq s} \bigg(\bigg|D^\al\bigg(f-\sum_{n=1}^{m+1}g_n\bigg)\bigg|+\big|D^\al g_{m+2}\big| +  \bigg|D^\al\bigg(\sum_{n=m+3}^\infty g_n\bigg)\bigg|\bigg)\\
  &<\si\ms\big[(\ep_{m+1}+2\ep_m) +(\ep_{m+2}+\ep_{m+1}) +\sum_{n=m+3}^\infty(\ep_n+\ep_{n-1})\big]\\
  &<2\si\ms (\ep_m+2\ep_{m+1})<\min_{x\in S\cap (K_{m+1}\minus K_m)}\tilde\ep(x)\,,
\end{align*}
where $\si$ was the number of multiindices $\al$ with $|\al|\leq s$. The better-than-uniform approximation lemma then follows.
\end{proof}

Let us now complete the proof of the Approximation Theorem~\ref{T.Carleman}. As the local Beltrami field $v$ satisfies $\Div v=0$ in the set $S$, acting with the operator $\rot+\la$ on the equation $\rot v=\la v$ we readily derive that $(\De+\la^2)v=0$, i.e., that each component $v_i$ of the vector field (in Cartesian coordinates) satisfies the scalar equation $(\De+\la^2)v_i=0$. By Lemma~\ref{L.Carleman} above, we can then approximate the function $v_i$ by a global solution $w_i$ of the equation $(\De+\la^2)w_i=0$ in the $C^s$ better-than-uniform sense. To be more concrete, for later convenience we shall take $s=r+2$ (with $r$ the order of approximation in Theorem~\ref{T.Carleman}) and choose an error function $\tilde\ep(x)$ to be specified later, thereby assuming that the global solution $w_i$ approximates the function $v_i$ pointwise in the set $S$ as
\[
\sum_{|\al|\leq r+2}\big|D^\al v_i(x)-D^\al w_i(x)\big|<\tilde\ep(x)\,,\qquad x\in S\,.
\]

Let us now denote by $w$ the vector field in $\RR^3$ whose Cartesian components are $w_i$. Defining the auxiliary vector field $\tw:=\rot w/\la$, an easy computation using the above $C^{r+2}$ estimate for the difference $v_i-w_i$ yields that the difference of the fields $v$ and $\tw$ satisfies the following $C^{r+1}$ pointwise bound in the set $S$:
\begin{equation}\label{vhv}
  \sum_{|\al|\leq r+1}\big|D^\al v(x)-D^\al\tw(x)\big|=\frac1\la\sum_{|\al|\leq r+1}\big|D^\al\rot v(x) -D^\al\rot w(x)\big|<\frac{6\ms\tilde\ep(x)}\la\,.
\end{equation}
By construction, the vector field $\tw$ obviously satisfies the equations $(\De+\la^2)\tw=0$ and $\Div\tw=0$, which in turn implies that
\[
(\rot+\la)(\rot-\la)\tw=0
\]
because, being divergence-free, the Laplacian of this field is $\De \tw=-\rot\rot \tw$. Hence the analytic vector field $\tv:=(\rot+\la)\tw/(2\la)$ satisfies the Beltrami equation $\rot\tv=\la\tv$ and, as a consequence of the estimate~\eqref{vhv}, the pointwise $C^r$ bound
\[
\sum_{|\al|\leq r}\big|D^\al\tv(x)-D^\al v(x)\big| =\frac1{2\la}\sum_{|\al|\leq r}\big|D^\al(\rot+\la)(\tw(x)-v(x))\big|<\frac{(6+3\la)\tilde\ep(x)}{\la^2}
\]
in the set $S$. Thus $\tv$ is the field whose existence was claimed in the theorem, provided that the error function $\tilde\ep(x)$ is chosen small enough (e.g., smaller than $\la^2\ep(x)/(6+3\la)$, where $\ep(x)$ is the error function in the statement of Theorem~\ref{T.Carleman}). 

\begin{remark}
It should be noticed that we have used that the parameter $\la$ is nonzero in the last part of the proof; in fact, if $\la=0$ one cannot hope to approximate the local solutions by global ones because an irrotational vector field defined in a simply connected region cannot have a periodic trajectory.
\end{remark}

\section{Applications and examples}
\label{S.examples}

To illustrate the Main Theorem~\ref{T.main}, we shall address a problem of Etnyre and Ghrist concerning the existence of steady solutions of the Euler equation having stream lines of all knot and link types. We shall also indicate how Theorem~\ref{T.main} can be extended to any open analytic Riemannian $3$-manifold.

\begin{figure}[!t]
\centering
\includegraphics[scale=0.3,angle=0]{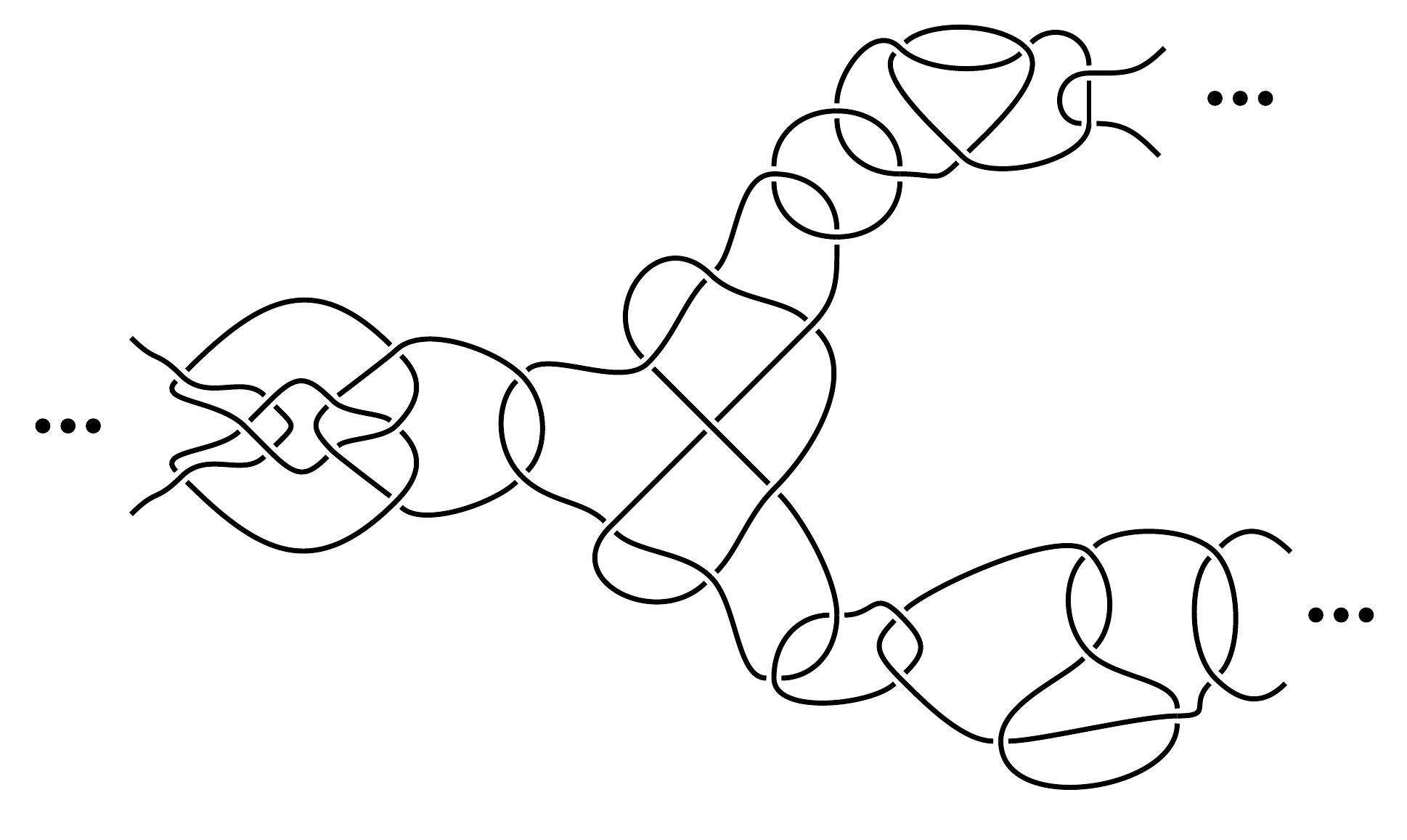}
\caption{Example of a locally finite link containing the Borromean
  rings, the trefoil, the figure eight and the $(7,4)$ knots.} 
\label{knot}
\end{figure}

\begin{example}\label{E.EG}
While in the literature one can explicitly find questions regarding the existence of solutions whose stream or vortex lines are simple links (e.g., the Borromean rings in~\cite{Mo85}), a particularly attractive question on the complexity of steady Euler flows was formulated by Etnyre and Ghrist in~\cite{EG00}: Are there any steady solutions to the Euler equation having stream lines of all  knot and link types in Euclidean $3$-space?

Theorem~\ref{T.main} readily provides a positive answer to Etnyre and Ghrist's question for locally finite links. To see this, it suffices to notice that there is a countable number of isotopy classes of tame knots and locally finite links in $\RR^3$~\cite{Ro90}, which allows us to construct a locally finite link $L$ containing a sublink of each isotopy class by the noncompactness of $\RR^3$ (cf.\ Figure~\ref{knot} for an example of this kind of construction). 
\end{example}

\begin{example}\label{E.Riem}
  A straightforward modification of the proof of Theorem~\ref{T.main} actually works for the Euler equation in a general Riemannian open $3$-manifold $M$ of class~$C^\om$. Let us briefly elaborate on this point. The Euler equation automatically makes sense on a Riemannian manifold, with $\nabla$ being the covariant derivative associated to the metric $g$ and $u$ a vector field in $M$. A Beltrami field is most easily characterized in this context in terms of the associated $1$-form $\be$, which satisfies the equation $\star d\be=\la\be$.
  
  Mutatis mutandis, both the statements and proofs in the Steps~1 to~3 of the proof of Theorem~\ref{T.main} remain valid in this more general situation. This is not surprising, as they correspond to a (rather delicate) local construction which does not make use of any particular feature of the Euclidean metric.

  Step~4 is also valid after minor modifications. In the non-Euclidean setting, it is convenient to work directly with the $1$-form $\be$, so that the role of the Euclidean Laplacian in Section~\ref{S.global} is now played by the Hodge Laplacian $\De:=-(dd^*+d^*d)$. It is evident that better-than-uniform approximation of vector fields can be painlessly translated into better-than-uniform approximation of the associated $1$-forms (generally with different error functions). The $1$-form $\be$ then satisfies the equation $(\De+\la^2)\be=0$, for which Lemma~\ref{L.Carleman} and its proof can be easily adapted. The remaining parts of Step~4 remain unchanged, and so does Step~5. The analog of Theorem~\ref{T.main} on $M$ thus follows.

\end{example}

\section*{Acknowledgments}

The authors are grateful to an anonymous referee for valuable suggestions that improved the  presentation of the article. The authors are indebted to Javier Rodríguez-Laguna for producing Figure~\ref{knot}. The source code of the programme he designed for drawing finite links can be found at \texttt{http://saraswati.uc3m.es/xknots/}. The second author thanks the ETH Zürich for hospitality and support. This work is supported in part by the MICINN under grants no.\ FIS2008-00209 (A.E.) and MTM2007-62478 (D.P.-S.) and by Banco Santander--UCM under grant no.\ GR58/08-910556 (A.E.). The authors acknowledge the MICINN's financial support through the Ramón y Cajal program.


\begin{thebibliography}{00}\frenchspacing
  
  
\bibitem{Ar65}
  V.I. Arnold, Sur la géométrie différentielle des groupes de Lie de dimension infinie et ses applications à l'hydrodynamique des fluides parfaits, Ann. Inst. Fourier 16 (1966) 319--361. 

\bibitem{BG88}
T. Bagby, P.M. Gauthier, Approximation by harmonic functions on closed subsets of Riemann surfaces, J. Anal. Math. 51 (1988) 259--284.

\bibitem{BR96}
M.A. Berger, R.L. Ricca, Topological ideas and fluid mechanics, Phys. Today 49 (1996) 28--34. 
  
\bibitem{BCR98}
J.  Bochnak, M. Coste, M.F. Roy, {\em Real algebraic geometry}, Springer-Verlag, Berlin, 1998.


\bibitem{EG99}
  J. Etnyre, R. Ghrist, Stratified integrals and unknots in inviscid flows, Contemp. Math. 246 (1999) 99--111.
  
\bibitem{EG00}
J.   Etnyre, R. Ghrist, Contact topology and hydrodynamics III. Knotted orbits, Trans. Amer. Math. Soc. 352 (2000) 5781--5794.

\bibitem{FH91}
M.H. Freedman, Z.X. He, Divergence-free fields: energy and asymptotic crossing number, Ann. of Math. 134 (1991) 189--229.

\bibitem{Ha82}
P. Hartman, {\em Ordinary differential equations}, Birkhäuser, Boston, 1982.

\bibitem{He66}
M. Hénon, Sur la topologie des lignes de courant dans un cas particulier, C. R. Acad. Sci. Paris 262 (1966) 312-314.

\bibitem{Hi76}
M.W. Hirsch, {\em Differential topology}, Springer-Verlag, New York,
1976.


\bibitem{HPS77}
M.W. Hirsch, C.C. Pugh, M. Shub, Invariant Manifolds, Lecture Notes in Mathematics 583, Springer-Verlag, New York, 1977.

\bibitem{Kh05}
  B. Khesin, Topological fluid dynamics, Notices AMS 52 (2005) 9--19.

\bibitem{Acta}
  G. Koch, N. Nadirashvili, G.A. Seregin, V. Sverák, Liouville theorems for the Navier--Stokes equations and applications, Acta Math. 203 (2009) 83--105.

\bibitem{KP81}
  S.G. Krantz, H.R. Parks, Distance to $C^k$ hypersurfaces, J. Differential Equations 40 (1981) 116--120.

\bibitem{La56}
  P.D. Lax, A stability theorem for solutions of abstract differential equations, and its application to the study of the local behavior of solutions of elliptic equations, Comm. Pure Appl. Math. 9 (1956) 747--766.

\bibitem{LS00}
  P. Laurence, E.W. Stredulinsky, Two-dimensional magnetohydrodynamic equilibria with prescribed topology, Comm. Pure Appl. Math. 53 (2000) 1177--1200.

\bibitem{MB02}
A.J.    Majda, A.L. Bertozzi, {\em Vorticity and incompressible flow}, Cambridge University Press, Cambridge, 2002. 

\bibitem{Ma56}
B.  Malgrange, Existence et approximation des solutions des équations aux dérivées partielles et des équations de convolution, Ann. Inst. Fourier 6 (1955--1956) 271--355.

\bibitem{Ma59}
  W.S.   Massey, On the normal bundle of a sphere imbedded in Euclidean space, Proc. Amer. Math. Soc. 10 (1959) 959--964.

\bibitem{Mo85}
  H.K.  Moffatt, Magnetostatic equilibria and analogous Euler flows of arbitrarily complex topology. I. Fundamentals, J. Fluid Mech. 159 (1985) 359--378; II. Stability considerations, ibid. 166 (1986) 359--378.

\bibitem{Ro90}
D. Rolfsen, {\em Knots and links}, Publish or Perish, Houston, 1990.



\bibitem{Vo03}
T. Vogel, On the asymptotic linking number,
Proc. Amer. Math. Soc. 131 (2003) 2289--2297.


  
\end{thebibliography}
\end{document}